\documentclass{procpavia}

\usepackage{amsmath}
\usepackage{amssymb}

\usepackage{graphicx}
\usepackage{epsfig}
\usepackage{rotating}

\usepackage{dcolumn}
\usepackage{bm}

 \def \bm #1{ {\mathbf #1} }
\def\beq{\begin{equation}}
\def\eeq{\end{equation}}

\begin{document}
\pagestyle{prochead}

   \title{
 DEUTERON DISTRIBUTION IN NUCLEI AND IN CORRELATED NUCLEAR MATTER
}
     \author{Alexei Yu.~Illarionov}
\email{Alexei.Illarionov@jinr.ru}
	  \affiliation{Joint Institute for Nuclear Research,
		 141980 Dubna, Moscow Region, Russia}

\begin{abstract}
We compute the distribution of quasideuterons in 
doubly closed shell nuclei and infinite correlated nuclear matter.
The ground states of $^{16}$O and $^{40}$Ca are described in $ls$ coupling
using a realistic hamiltonian including 
the Argonne $v_{8}^\prime$ and the Urbana IX models of 
two-- and three--nucleon potentials, respectively. The  
nuclear wave function contains central and tensor correlations, 
and correlated basis functions theory is used to evaluate the 
distribution of neutron-proton pairs, 
having the deuteron quantum numbers, as a function of their total 
momentum. By computing the number of deuteron--like pairs we 
are able to extract the Levinger's factor and compare to both the 
available experimental data and the predictions of the local 
density approximation, based on nuclear matter estimates. 
The agreement with the experiments is excellent, whereas the 
local density approximation is shown to sizably overestimate the 
Levinger's factor in the region of the medium nuclei.
\end{abstract}
\maketitle
\setcounter{page}{1}

\section{Introduction}
 \label{sec:Introduction}

Within the Levinger's \emph{quasideuteron} (QD) model \cite{LKG}
the nuclear photoabsorption cross section $\sigma_A(E_\gamma)$, 
above the giant dipole resonance and below pion threshold, 
is assumed to be proportional to the break-up cross section  
of a deuteron in hadronic matter, $\sigma_{QD}(E_\gamma)$:  
\begin{equation} 
\sigma_A(E_\gamma) = {\mathcal P_D} \  
\sigma_{QD}(E_\gamma)\ ,
\label{levinger:formula}
\end{equation}
where $E_\gamma$ is the photon energy and $\mathcal P_D$ is interpreted
as the effective number of the nucleon--nucleon (NN) pairs of the QD type
(see \cite{Nemetz:1988} and references therein).
$\mathcal P_D$ is written in the form
\begin{equation}
{\mathcal P_D} = {\mathrm L}\ \left[ \dfrac{Z({A - Z})}{A} \right]\ , 
\label{def:L}
\end{equation}
where $A$ and $Z$ are the mass and atomic numbers of the nucleus  
and $\mathrm L$ is the so called Levinger's factor.  
$\mathcal P_D$ can be calculated for a given nuclear ground state 
wave function, thus allowing for a \emph{microscopic} interpretation of 
the \emph{phenomenological} Levinger's factor.

The value of $\mathrm L$ has been extracted from experiments according to 
the following two models: i) the Levinger's model \cite{Levinger:1979}, in
which $\sigma_{QD}(E_\gamma)$ 
is taken as the deuteron cross section damped by an exponential function, 
taking care of Pauli blocking of the final states available to the nucleon
ejected from the QD, and ii) the Laget's model \cite{Laget:1981}, 
which associates $\sigma_{QD}(E_\gamma)$ with the transition 
amplitudes of virtual ($\pi\ +\ \rho$)--meson exchanges between 
the two nucleons of the QD pair.

Both models provide satisfactory fits of photoreaction data in heavy 
nuclei, but yield different values of the Levinger's factor, 
${\mathrm L}_\text{Lev}(A)$ 
and ${\mathrm L}_\text{Laget}(A)$, ${\mathrm L}_\text{Laget}(A)$ being 
$\sim 20\%$ larger than ${\mathrm L}_\text{Lev}(A)$.

The effective number of deuteron-like pairs, as well as of three- and
four-body structures, in spherical nuclei has been investigated 
within the shell model approach in Refs.~\cite{kadm:1981,vals:1981}.  
In a recent paper \cite{Benhar:2002} (referred to as I hereafter) we have
analyzed the properties of deuteron-like structures in infinite symmetric
nuclear matter (NM), described by a hamiltonian containing the realistic
Urbana $v_{14}$ NN potential and the Urbana TNI many-body potential
\cite{Lagaris:1981}. 
A correlated wave function having spin-isospin dependent, central and 
tensor correlations has been used within the correlated basis 
functions (CBF) theory to compute the QD distribution function in matter 
and extract the NM Levinger's factor at equilibrium density, 
${\mathrm L}_{NM}=11.63$, to be compared to the
empirical estimate, ${\mathrm L}_\text{expt}(A=\infty)=9.26$.

CBF theory has established itself as one of the most effective tools 
to realistically study, from a microscopic viewpoint, properties 
of infinite matter of nucleons  ranging from the equation of state
\cite{Wiringa:1988,Akmal:1998} to the momentum distribution
\cite{Fantoni:1984} and the one-- and two--body Green's functions
\cite{Benhar:1989-2000}.
In the last decade these studies have been successfully extended to 
deal with finite nuclei
\cite{Co:1992-1996,Fabrocini:2000,Fabrocini:2001}.

In the paper \cite{Benhar:2003} (referred to as II hereafter) we extended the
CBF many body approach, used in I for NM, to evaluate \emph{ab initio} the
momentum distribution,  $P_D({\bm k}_D)$, and the total number per particle,
${\mathcal P_D}/A$, of QD pairs in the doubly closed shell nuclei $^{16}$O
and $^{40}$Ca, described in the $ls$ coupling scheme. From ${\mathcal P_D}/A$ 
we then extracted the corresponding Levinger's factors. 

Following the approach developed in I and II, in a $A$--nucleon system 
the momentum distribution of QD pairs can be written as the overlap between
the nuclear and deuteron ground state wave functions in terms of the
\emph{natural orbits} (NO), $\phi_\alpha^{NO}(k)$ \cite{Fabrocini:2001},
and their occupation numbers, $n_{\alpha=(nlm)}$, which are obtained by
expanding the one--body density matrix in multipoles.
\begin{equation}
{\mathcal P_D}({\bm k}_D)=
\sum_{\alpha,\alpha^\prime} \, 
n_{\alpha} n_{\alpha^\prime} \, 
{\mathcal P}_{\mathcal D}^{\alpha,\alpha^\prime}({\bm k}_D) \, ,
\label{sum:PD_kD}
\end{equation}
with
\begin{equation}
{\mathcal P}_{\mathcal D}^{\alpha,\alpha^\prime}({\bm k}_D) =
\dfrac{\nu^2}{16} \, \dfrac{(2\pi)^3}{4\pi} \,
\left[
| \Psi^{\alpha,\alpha^\prime}_S({\bm k}_D)|^2 \, + 
\sum_{s=-2}^2 | \Psi^{\alpha,\alpha^\prime;\,s}_D({\bm k}_D)|^2 
\right] \, ,
\label{calc:PDkD}
\end{equation}
where,
\begin{gather}
\Psi^{\alpha,\alpha^\prime}_S({\bm k}_D)=
\int d^3k \,
 \phi_{\alpha}^{NO\dagger} \left( \dfrac{{\bm k}_D}{2} + {\bm k} \right)
  \, U(k) \,   
 \phi_{\alpha^\prime}^{NO} \left( \dfrac{{\bm k}_D}{2} - {\bm k} \right) 
\, ,
\label{def:Psi_S} \\
\Psi^{\alpha,\alpha^\prime;\,s}_D({\bm k}_D) = \sqrt{\dfrac{4\pi}{5}}
\int d^3k \,
 \phi_{\alpha}^{NO\dagger} \left( \dfrac{{\bm k}_D}{2} + {\bm k} \right)
  \, W(k) \,   
 \phi_{\alpha^\prime}^{NO} \left( \dfrac{{\bm k}_D}{2} - {\bm k} \right) \,   
 Y_{2s}(\widehat{{\bm k}})
\, , 
\label{def:Psi_D}
\end{gather}
and
\begin{equation}
\phi_{nlm}^{NO}({\bm q}) =  
\phi_{nl}^{NO}(q) \, Y_{lm}(\widehat{{\bm q}}) \, , 
\end{equation}
$Y_{lm}(\widehat{{\bm q}})$ being the spherical harmonics.

In the independent particle model (IPM), an $A$--body wave function
$\Psi_{A}(R)$ is the Slater determinant of single particle orbitals
$\phi_\alpha(i)$, which are eigenfunctions of a suitable single particle
hamiltonian, and $n_\alpha^{IPM}=1$, $\phi_\alpha^{NO} \equiv \phi_\alpha$
for occupied states, whereas $n_\alpha^{IPM}=0$ for unoccupied states.
For nuclear matter, the orbitals $\phi_\alpha(i)$ are plane waves
corresponding to a noninteracting Fermi gas of nucleons with momenta
$|{\bm k}|\le k_F = (6 \pi^2 \rho_{NM}/\nu)^{1/3}$,
$\nu=4$ and $\rho_{NM}$ are the NM spin--isospin degeneracy and
density, respectively. Deviations from IPM provide
a measure of correlation effects, as they allow higher NO to become
populated with $n_\alpha \neq 0$.

The $U(r)=u_D(r) - \Delta u(r)$ and $W(r)=w_D(r)-\Delta w(r)$ functions
account for the medium correlations effect on the bare components of the DWF,
$u_D(r)$ and $w_D(r)$. Their explicit expressions, in terms of the
6 central and non--central spin--isospin dependent correlation functions,
which are variationally fixed by minimizing the ground state energy
(see \cite{Fabrocini:2000} and references therein),
are given in I.

\section{Results}
\label{sec:Results}

Last generation NN potentials are able to fit deuteron properties
and the Nijmegen 93  nucleon--nucleon scattering 
phase-shifts \cite{Nijmegen:1993}  
up to the pion--production threshold ($\sim 4000$ data points) with a 
$\chi^2\sim 1$.
The Argonne $v_{18}$, belonging to this generation, is given by 
the sum of 14 isoscalar and 4 isovector terms, including 
charge-symmetry and charge-invariance breaking 
components \cite{Wiringa:1995}.  
In this work we have used  a simpler NN potential, referred to as
Argonne $v_{8}^\prime$, obtained from the the full Argonne $v_{18}$
retaining  only the first eight operatorial terms,
corresponding the central and non-central spin-isospin dependent components 
plus spin--orbit and spin--orbit/isospin components.
The Argonne  $v_{8}^\prime$ is constructed
in such a way to reproduce the isoscalar part of the full $v_{18}$ in the 
$S$, $P$ and $^3D_1$ waves and the $^3D_1$--$^3S_1$ coupling.
The $v_{8}^\prime$ parameterization, while allowing for a fully
realistic NN interaction, makes the use of modern
many-body methods much more practical.
 It has been found that the differences between
Argonne $v_{8}^\prime$ and the full $v_{18}$ contribute very little to the 
binding
energy of light nuclei and nuclear matter, and can be safely estimated
either by perturbation theory or from FHNC/SOC calculations. 

It is well known that, to quantitatively describe the properties of nuclei 
with $A>2$, modern NN interactions need to be supplemented with 
three-body forces. 
The Urbana IX (UIX) model provides a very good description 
of the energies of both the ground and the low-lying 
excited states of light nuclei ($A \le 8$).
In the present calculations we use Argonne $v_{8}^\prime$ + 
UIX interaction,
which will be referred to as as the AU8$^\prime$ model. 
This interaction has
already been used in the variational FHNC/SOC calculations of
Ref.~\cite{Fabrocini:2001} as well as in the quantum 
Monte Carlo simulations of Ref.~\cite{Pudliner:1997}.

For the single particle wave functions, $\phi_\alpha(i)$, 
entering the shell model wave function $\Phi_0$, we have used the solution 
of the single particle Schr\"odinger equation with a Woods--Saxon
potential,
\begin{equation}
V_{WS}(r)= \dfrac{V_0}{1+\exp{[(r-R_0)/a_0]}} \, .
\label{WS}
\end{equation}
In principle, the parameters of the correlation functions, $f_p(r)$, and 
of the Woods--Saxon potential may be both fixed by 
minimizing the ground state energy.
This complete minimization was performed for the AU8$^\prime$ model 
in Ref.~\cite{Fabrocini:2000}, and provided a binding energy per nucleon
of $B/A = 5.48~\text{MeV}$ in $^{16}$O and $B/A = 6.97~\text{MeV}$ in
$^{40}$Ca (the experimental values are $7.97~\text{MeV}$ in $^{16}$O and
$8.55~\text{MeV}$ in $^{40}$Ca).
These differences are compatible with the results of nuclear 
matter calculations at saturation density, $\rho_{NM} = 0.16~\text{fm}^{-3}$, 
carried out with the same hamiltonian. In fact, the FHNC/SOC
nuclear matter energy per nucleon is $E_{NM}/A=-10.9~\text{MeV}$
\cite{Fabrocini:2000}, to be compared to the empirical value of 
$-16~\text{MeV}$.

However, the calculated root mean square radii of the two nuclei 
turned out to be
$R = 2.83~\text{fm}$ in $^{16}$O and $R = 3.66~\text{fm}$ in $^{40}$Ca,
showing a difference of $\sim 5\%$  with the experimental values, 
$R_\text{expt} = 2.73~\text{fm}$ and $R_\text{expt} = 3.48~\text{fm}$,
respectively. Moreover, the one--body densities were not in close 
agreement with the experimental ones. In order to take care of this 
feature of the variational approach, a set of single particle wave 
functions providing an accurate description
of the empirical densities was chosen, and the energy was then minimized
with respect to the correlation functions only. The resulting radii were
$R = 2.67~\text{fm}$ ($^{16}$O) and $R = 3.39~\text{fm}$ ($^{40}$Ca), with
a density description very much improved. The energies obtained by this
partial minimization procedure were $B/A = 5.41~\text{MeV}$ in $^{16}$O
and $B/A = 6.64~\text{MeV}$ in $^{40}$Ca, largely within the accuracy of
the FHNC/SOC scheme. Here, we adopt this same wave function,
whose parameters are given in Table~V of Ref.~\cite{Fabrocini:2000}.

The structure of the NO in $^{16}$O and $^{40}$Ca is discussed at length 
Ref.~\cite{Fabrocini:2001}.
 Here we limit ourselves to recall some of their main characteristics. 
The effect of correlations is mostly visible in the $1s$ orbital, 
where the NO are larger than the shell model ones at short distances,
resulting in stronger localization. The influence on the shape of the other 
occupied
shell model orbitals is negligible.
The occupation of the NO corresponding to the fully occupied shell model 
states is depleted by  $9.6\%$ in $^{16}$O and by $10.5\%$ in $^{40}$Ca, 
with a maximum depletion of $\sim 22\%$ for the $2s$ state in $^{40}$Ca. 
As a consequence, the lowest mean field unoccupied states become sizably 
populated ($n_{2s} \approx n_{2p} \approx n_{1d} \approx 0.02$ in $^{16}$O 
and $n_{3s} \approx 0.05$, $n_{2p} \approx 0.02$, $n_{2d} \approx 0.03$
in $^{40}$Ca). 
These two effects are largely due to the presence of the tensor correlation.  

\begin{figure}[t]
{\epsfig{figure=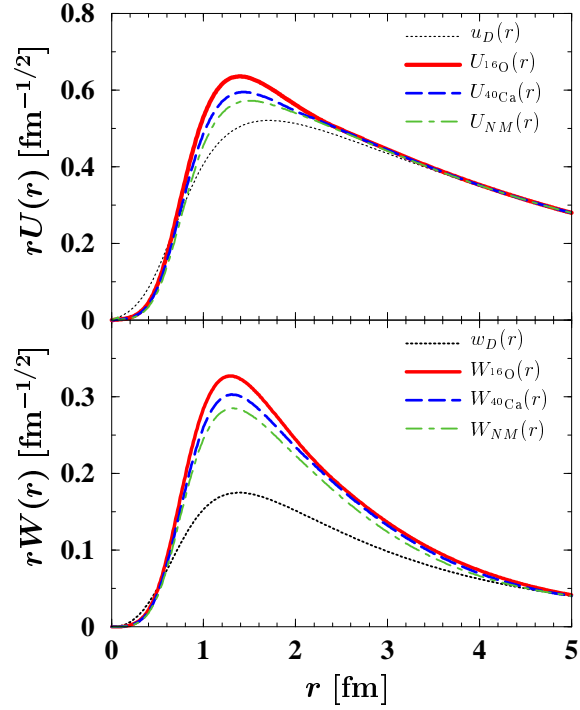, width=7.5cm}}
\caption{%
Radial components $U(r)$ and $W(r)$ of 
the AU8$^\prime$ QD wave functions
in $^{16}$O, $^{40}$Ca and nuclear matter.
Upper panel: the solid and dashed lines show the radial dependence of
$U_A(r)$ for $^{16}$O and $^{40}$Ca, respectively. 
The dot-dashed and dotted lines correspond 
to the nuclear matter $U_{NM}(r)$ and the bare $u_D(r)$.
Lower panel: as in the upper panel for the $d$--wave components
of the QD and deuteron wave functions.
}%
\label{Fig:UW-r}
\end{figure}

Fig.~\ref{Fig:UW-r} shows the behavior of $U(r)$ and $W(r)$ in
$^{16}$O, $^{40}$Ca and nuclear matter, evaluated using the hamiltonian
AU8$^\prime$. For comparison, we also show the bare components of the
Argonne $v_{8}^\prime$ DWF. It appears that the main differences between
deuteron and QD occur at $r \lesssim 2~\text{fm}$. 
At small relative distances both  $U(r)$ ($r \lesssim 1~\text{fm}$)
and $W(r)$ ($r \lesssim 0.5~\text{fm}$) are slightly suppressed 
with respect to $u_D(r)$ and  $w_D(r)$. On the contrary, they are 
appreciably enhanced at larger distances. These effects are 
more visible for the lightest nucleus.

\begin{figure}[b]
\vspace{-0.3cm}
{\epsfig{figure=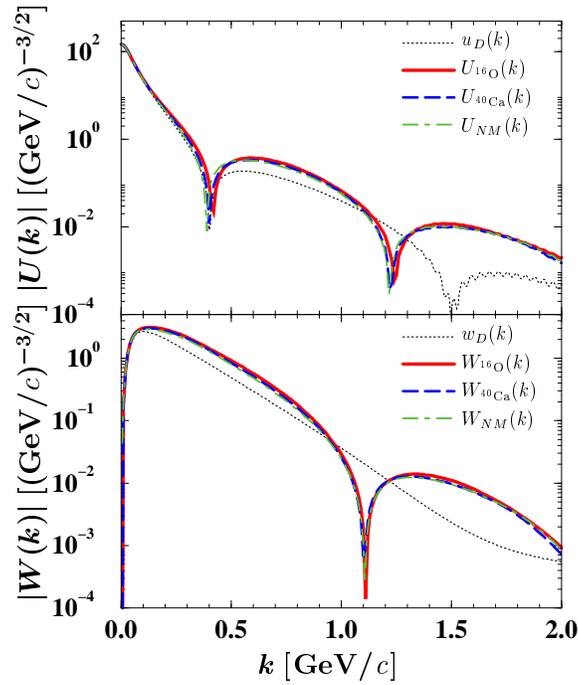, width=7.5cm}}
\caption{%
As in Fig.~\protect\ref{Fig:UW-r} in momentum space.
}%
\label{Fig:UW-q}
\end{figure}

The differences between nuclear matter and nuclei mostly disappear 
in the Fourier transforms, $|U(k)|$, $|W(k)|$, $|u_D(k)|$ and
$|w_D(k)|$, whose behavior is displayed in Fig.~\ref{Fig:UW-q}. 
The nuclear medium shifts the second minimum of $|u_D(k)|$ towards 
lower values of k, as obtained in I for nuclear matter with the 
Urbana $v_{14}$ potential. 
The Argonne $v_{8}^\prime$ $|w_D(k)|$ does not exihibit any diffraction 
minimum, which, however, appears in  $|W(k)|$. 

\begin{figure}[t]
{\epsfig{figure=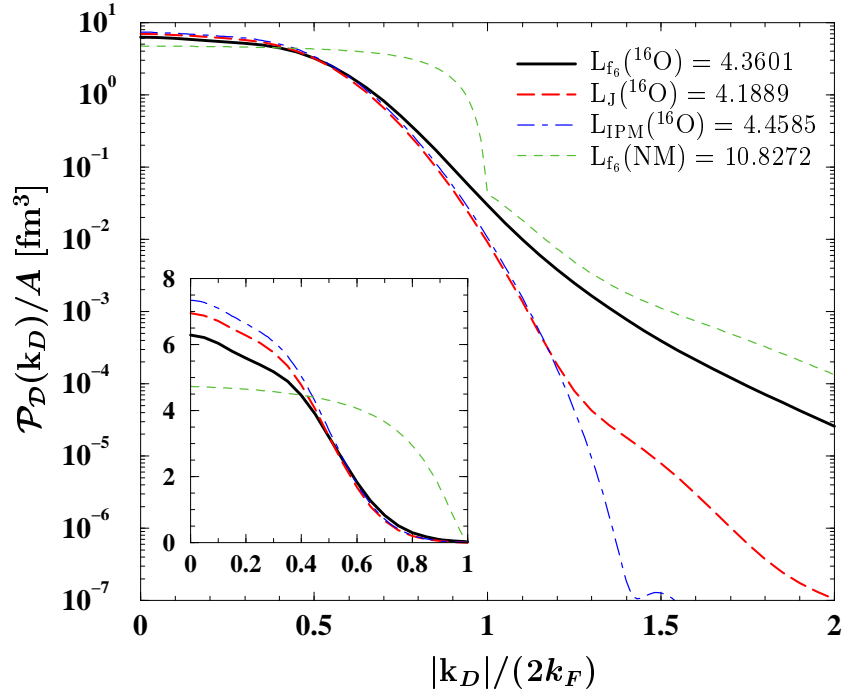, width=11cm}}
\caption{%
Momentum distribution of QD pairs in $^{16}$O as a function of the 
total momentum $|{\bm k}_D|$ (see Eq.~(\protect\ref{sum:PD_kD})).
The solid, dashed and dash--dotted lines are the results obtained 
within the $f_6$ and Jastrow  correlation models and IPM, respectively. 
The short--dashed line displays
the $f_6$ momentum distribution of the QD in nuclear matter at
equilibrium density, $\rho_{NM} = 0.16~\text{fm}^{-3}$.
The insert shows a blow up of the region $|{\bm k}_D|/(2k_F) <$ 1,
plotted in linear scale. The Levinger factors, ${\textrm L}(A)$, 
for the various calculations are also reported.
}%
\label{Fig:PD-kD-16O}
\end{figure}

\begin{figure}
\vspace{-0.3cm}
{\epsfig{figure=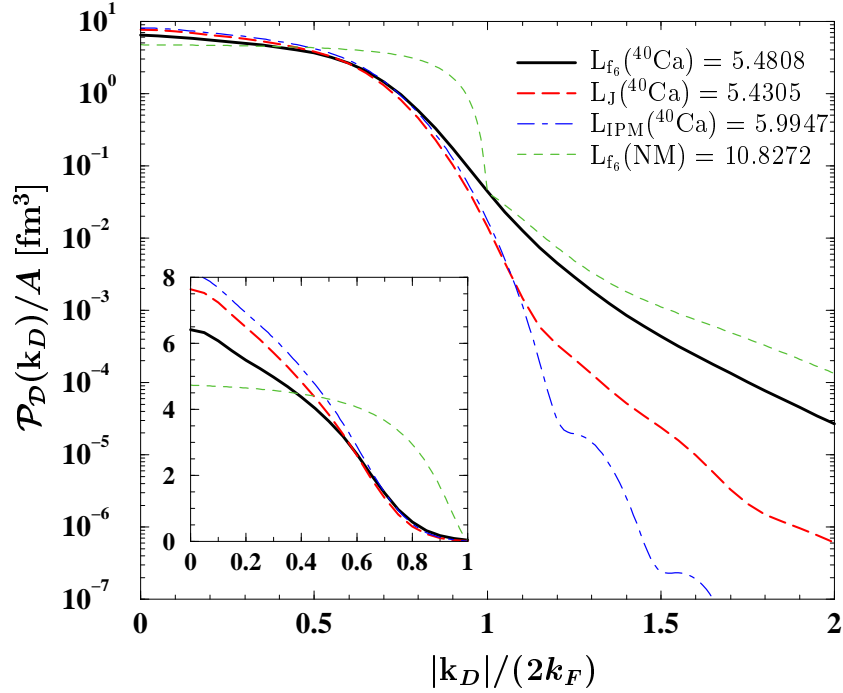, width=11cm}}
\caption{%
 As in Fig.~\protect\ref{Fig:PD-kD-16O} for $^{40}$Ca.
}%
\label{Fig:PD-kD-40Ca}
\end{figure}

The distribution of deuteron pairs with total momentum ${\bm k}_D$, 
${\mathcal P_D}({\bm k}_D)$, resulting from our approach is displayed
by the solid line in Fig.~\ref{Fig:PD-kD-16O} for $^{16}$O and in
Fig.~\ref{Fig:PD-kD-40Ca} for $^{40}$Ca. The following comments are in order:
\begin{itemize}
\item[(i)] $NN$ correlations introduce high momentum components in the
distribution. The full ${\mathcal P_D}({\bm k}_D)$ is
strongly enhanced with respect to ${\mathcal P}_{\mathcal D}^{IPM}({\bm k}_D)$ 
at large $|{\bm k}_D|$, and it is correspondingly depleted at small
$|{\bm k}_D|$. The depletion is mostly due to the non--central
tensor correlations.
\item[(ii)] The effect of state--dependent correlations is large, as one can
see by comparing the full ${\mathcal P_D}({\bm k}_D)$ with the Jastrow
model ${\mathcal P}_{\mathcal D}^J({\bm k}_D)$ (obtained by retaining
only the scalar component in the two--body correlation operator).
\item[(iii)] The tail of ${\mathcal P_D}({\bm k}_D)$ is appreciably 
different from that of nuclear matter. At $|{\bm k}_D| = 4k_F$ the 
difference is still a factor $\sim 10$ for both $^{16}$O and $^{40}$Ca.
\end{itemize}

\begin{figure}[t]
{\epsfig{figure=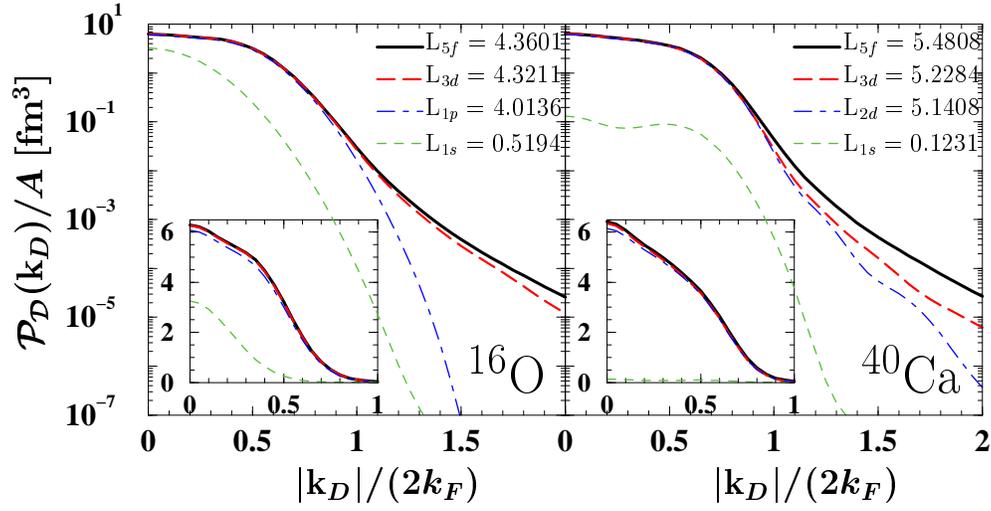, width=13cm}}
\caption{%
Convergence of ${\mathcal P_D}({\bm k}_D)/A$ for $^{16}$O and $^{40}$Ca
in the number of natural orbits included in the summation of
Eq.~(\ref{sum:PD_kD}). The results have been obtained within
the $f_6$ correlation model.
}%
\label{Fig:PD-kD-c}
\end{figure}

Fig.~\ref{Fig:PD-kD-c} displays the convergence of ${\mathcal P_D}({\bm k}_D)$
in the number of natural orbits included in the sum of
Eq.~(\ref{sum:PD_kD}). Full convergence is reached with the inclusion
of orbits up to $5f$ for both $^{16}$O and $^{40}$Ca. The figure shows that,
in the case of $^{40}$Ca, the tail of ${\mathcal P_D}({\bm k}_D)$ is still
$\sim 10$ times too small if only orbitals up to $3d$ are included.

The total number of pairs of the QD type in both finite nuclei and
nuclear matter, $\mathcal P_D$ is obtained by integration of
${\mathcal P_D}({\bm k}_D)$ over ${\bm k}_D$:
\begin{equation}
\dfrac{\mathcal P_D}{A} = 3 \int \dfrac{d^3k_D}{(2\pi)^3} \
 \dfrac{{\mathcal P_D}({\bm k}_D)}{A} \ ,
\label{def:totprob}
\end{equation}
where the factor $3$ in the r.h.s corresponds to the spin multiplicity
of the deuteron, $2J_D+1$.

We have repeated the calculations for nuclear matter by using the AU8$^\prime$
interaction. The result ${\mathcal P_D}(NM)/A = 2.707$
(the corresponding Fermi gas model result is $3.382$) should be
compared to the value $2.895$, obtained in I with the Urbana $v_{14}$
two--nucleon plus the Urbana TNI many--body forces \cite{Lagaris:1981}
(which will referred to as the UU14 model). The corresponding numbers for
$^{16}$O and $^{40}$Ca turn out to be much smaller:
${\mathcal P_D}(^{16}\text{O})/A = 1.090$ and
${\mathcal P_D}(^{40}\text{Ca})/A = 1.370$, respectively.

The Levinger factor is easily obtained from ${\mathcal P_D}/A$ by means of 
Eq.~(\ref{def:L}). As we are dealing with symmetric matter ($N=Z=A/2$),
 ${\mathrm L}(A) = 4 \, {\mathcal P_D}/A$. Our estimates
$\textrm{L}_{\text{f}_6}(^{16}\text{O})$,
$\textrm{L}_{\text{f}_6}(^{40}\text{Ca})$ and $\textrm{L}_{\text{f}_6}(NM)$
for $^{16}$O, $^{40}$Ca and nuclear matter, corresponding to the $f_6$
correlation model, are reported in Figs.~\ref{Fig:PD-kD-16O} and
\ref{Fig:PD-kD-40Ca}. These results are not too different from the
values obtained within the independent particle and Jastrow models. 
This fact actually implies 
that the high momentum tail of ${\mathcal P_D}({\bm k}_D)$ is not 
relevant for the calculation of the Levinger factor $\textrm{L}$.
It has to be stressed that the Jastrow model turns out to 
consistently underestimate the Levinger factor.
The spatial structure of $np$ pairs having the deuteron 
quantum numbers has been investigated in Ref.~\cite{Forest:1996} 
in light (A=3,4,6 and 7) nuclei and $^{16}$O using a variational   
Monte Carlo approach and the Argonne $v_{18}$ two--nucleon 
and Urbana IX three--nucleon potentials. The estimated Levinger 
factor for $^{16}$O is 
$\textrm{L}_{\text{VMC}}(^{16}\text{O})=4.70$, 
comfortably close to our value,
$\textrm{L}_{\text{f}_6}(^{16}\text{O})=4.36$.

\begin{figure}[b]
\vspace{-0.2cm}
{\epsfig{figure=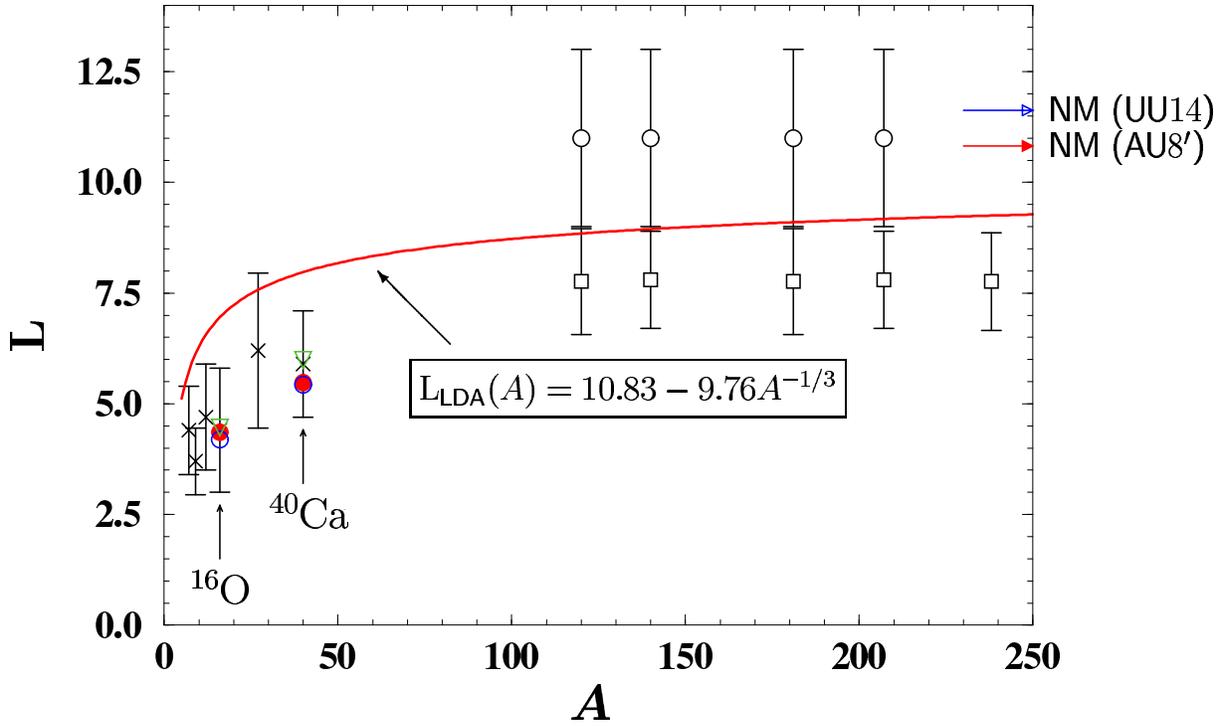, width=16cm}}
\caption{%
Levinger's factor ${\mathrm L}(A)$ for $^{16}$O,
$^{40}$Ca and nuclear matter (shown by the arrows for the UU14 and
AU8$^\prime$ forces).
The filled circles, the empty circles and the triangles 
show the Levinger's factors obtained within the $f_6$ and 
Jastrow correlation models and the IPM, respectively.
The LDA, as discussed in the text, is also reported (solid line).
The phenomenological values of ${\mathrm L}_\text{Lev}(A)$ 
corresponding to the photoreaction data of Lepretre \textit{et al.} 
\protect\cite{Lepretre:1981} (squares) and Ahrens 
\textit{et al.} \protect\cite{Ahrens:1975}
(crosses and diamonds) are taken from 
Ref.~\protect\cite{Anghinolfi:1986}. 
The empirical values of ${\mathrm L}_\text{Laget}(A)$, 
represented by circles in the heavy nuclei region, 
are from Ref.~\protect\cite{Carlos:1982}.
}%
\label{Fig:PD-Lev}
\end{figure}

Our results for the Levinger factors are summarized in Fig.~\ref{Fig:PD-Lev},
where they are also compared with the available experimental estimates.
The agreement with the photoreaction data of Ahrens \textit{et al.}
\cite{Ahrens:1975} for the case of $^{16}$O and $^{40}$Ca is rather
impressive. The ``experimental'' value,
${\mathrm L}_\text{expt}(\infty) = 9.26$, deduced from the phenomenological
formula
\begin{equation}
{\mathrm L}_\text{expt}(A) = 13.82 \dfrac{A}{R^3[\text{fm}^3]} \ ,
\label{expt:L}
\end{equation}
reported in Ref.~\cite{Anghinolfi:1986}, is $\sim 15\%$ smaller than our
theoretical value. In I, the surface contribution to ${\mathrm L}(A)$ has
been estimated exploiting the calculated enhancement factor
of the electric dipole sum rule for finite nuclei, 
$\mathcal K$, \cite{Fabrocini:1985}, obtained using CBF theory and LDA. 
The enhancement factor is related to experimental
data on photoreactions through the equation: 
\begin{equation}
1+{\mathcal K}_\text{expt} = 
\dfrac{1}{\sigma_0}\int_0^{m_\pi} \sigma_A(E_{\gamma}) dE_{\gamma} \ ,
\label{enhanc}
\end{equation}
where $\sigma_0 = 60\ [Z(A-Z)/A]$~MeV$\,mb$ and $m_\pi$ is the 
$\pi$--meson production threshold.
By using the same parameterization as in I for the surface term, we get:
\begin{equation}
{\mathrm L}_\text{LDA}(A) = 10.83 - 9.76 \ A^{-1/3} \ ,
\label{def:mass}
\end{equation}
for the AU8$^\prime$ interaction. ${\mathrm L}_\text{LDA}(A)$ is displayed
on Fig.~\ref{Fig:PD-Lev}. LDA turns out to be not satisfactory for
medium nuclei, such as $^{16}$O and $^{40}$Ca. Fig.~\ref{Fig:PD-Lev} also
report ${\mathrm L}_\text{Lev}(A)$ and ${\mathrm L}_\text{Laget}(A)$, as
extracted \cite{Anghinolfi:1986,Carlos:1982} from the available experimental
data on photoreactions. The computed Levinger's factors are almost
$A$--independent for heavy nuclei ($A > 100$), and result to be $\sim 15\%$
larger than ${\mathrm L}_\text{Lev}(A)$ and $\sim 25\%$ smaller than
${\mathrm L}_\text{Laget}(A)$. Such disagreement between theory and experiment
is likely to be ascribed to the sizable tail contributions to the electric 
dipole sum rule, absent in the definition of Eq.~(\ref{enhanc}).

\section{Conclusion}
\label{sec:Conclusion}

The Correlated Basis Function theory of the two--body density matrix has been
applied to microscopically compute the distribution of QD pairs 
carrying total momentum ${\bm k}_D$, ${\mathcal P_D}({\bm k}_D)$, in doubly
closed shell nuclei $^{16}$O and $^{40}$Ca and nuclear matter, 
starting from the realistic Argonne $v_8^\prime$ plus Urbana IX potential.

It has been found that $NN$ correlations produce a high momentum tail
in ${\mathcal P_D}({\bm k}_D)$ and, correspondingly a depletion at small
${\bm k}_D$ for both nuclei and nuclear matter. These effects are mainly due
to the presence of the state--dependent correlations associated with the
tensor component of the one pion exchange interaction.
Contrary to what happens for the one--nucleon momentum distibution, 
the tail of ${\mathcal P_D}({\bm k}_D)$
sizably differs from that of nuclear matter.

Summation of ${\mathcal P_D}({\bm k}_D)$ over ${\bm k}_D$ provides the total
number $\mathcal P_D$ of QD pairs, and, consequently, allows for an
\emph{ab initio} calculation of the Levinger's factor ${\mathrm L}(A)$.
The CBF result for nuclear matter is significantly reduced with respect to
the value obtained in I with the Urbana $v_{14}$ plus the Urbana TNI
many--body forces. The corresponding Levinger factors for  $^{16}$O
and $^{40}$Ca are much smaller than the nuclear matter value and in very good
agreement with the available photoreaction data analyzed within the  
quasideuteron phenomenology. In addition, our results show that LDA
overestimates ${\mathrm L}(A)$ in the region 
of the light--medium nuclei.

The ${\mathrm L}(A)$ resulting from the full calculation are relatively close 
to the corresponding
values obtained within the IPM and Jastrow models. Actually, the high
momentum tail of ${\mathcal P_D}({\bm k}_D)$  gives a small contributions
to the Levinger factor.
This feature indicates that the approximation used in our calculation 
(which amounts to including only diagrams at the dressed lowest order 
of the FHNC cluster expansion) is fully adequate.
However, it should be noticed that the Jastrow model underestimates 
the Levinger factor.

In addition, the analysis described in this paper shows that when a
deuteron is embedded in a nucleus, or in nuclear matter at equilibrium density,
its wave function gets appreciably modified by the surrounding medium. While 
in the case of the $S$-wave component the difference is mostly visible at
small  relative distance ($r < 1$ fm), the $D$-wave component of the QD
appears to be significantly enhanced, 
with respect to the deuteron $w_D(r)$,
over the range $r < 2~\text{fm}$. This effect is particularly evident
in the lightest nucleus.

\acknowledgments
The results presented in this contribution have been obtained in 
collaboration with Omar Benhar, Adelchi Fabrocini, Stefano Fantoni
and Gennadi I. Lykasov.
A.Yu.I. is grateful to the organizers of the Sixth Workshop on
Electromagnetically Induced Two-Hadron Emission in Pavia for the
kind invitation, financial support and fruitful atmosphere.

\newpage

\newpage

\newpage

\end{document}